\documentclass[aps,prl,10pt,twocolumn,superscriptaddress,showpacs,showkeys]{revtex4-1} 

\usepackage{graphicx}
\usepackage{hyperref}
\usepackage{setspace}
\usepackage{amsmath}
\usepackage{amssymb}
\usepackage{xcolor}
\usepackage{siunitx}
\usepackage[normalem]{ulem}

\renewcommand{\paragraph}[1]{\emph{#1}.---}

\begin{document}

\title{When elasticity affects drop coalescence}

\author{Pim J. Dekker}
\author{Michiel A. Hack}
\author{Walter Tewes}
\author{Charu Datt}
\author{Ambre Bouillant}
\author{Jacco H. Snoeijer}

\affiliation{Physics of Fluids Group, Mesa+ Institute, University of Twente, 7500 AE Enschede, The Netherlands}

\date{\today}

\begin{abstract}
The breakup and coalescence of drops are elementary topological transitions in interfacial flows. The breakup of a drop changes dramatically when polymers are added to the fluid. With the strong elongation of the polymers during the process, long threads connecting the two droplets appear prior to their eventual pinch-off. Here, we demonstrate how elasticity affects drop coalescence, the complement of the much studied drop pinch-off. We reveal the emergence of an elastic singularity, characterised by a diverging interface curvature at the point of coalescence. Intriguingly, while the polymers dictate the \emph{spatial} features of coalescence, they hardly affect the \emph{temporal} evolution of the bridge. These results are explained using a novel viscoelastic similarity analysis and are relevant for drops created in biofluids, coating sprays and inkjet printing.
\end{abstract}

\maketitle

Viscoelastic liquids are materials that can flow like ordinary liquids, yet respond elastically when excited by rapid deformations \citep{tanner2000engineering}. A prime example is provided by a ball of silly putty: it bounces like a rubber ball, but spreads out like a viscous liquid when left at rest on a table. Viscoelastic liquids are ubiquitous in biofluids and in technologies such as coating, printing, and polymer processing, and their flow poses many challenges. Of particular interest is how viscoelastic liquids behave near singularities \citep{renardy2000mathematical}, such as flows around sharp edges \citep{hinch1993, renardy1993, evans2008}, bubble cusps \citep{astarita1965, joseph1995, jimmy_annual2018} or during the breakup of drops \citep{ bazilevskii1990, entov1997jnnfm, anna2001, amarouchene2001prl, clasen2006,eggers-2020-jfm,bonn2020}. These flows involve regions of extreme polymer stretching which is why, for example, fluids containing polymers can produce long and stable threads during drop breakup \citep{anna2001} [see Fig.~\ref{fig:setup}(a)]. Liquid threads are indeed observed for biofluids such as saliva, where they play a role in the generation of aerosols \citep{abkarian2020}.

{In contrast to  breakup (see \citep{renardy2004self, eggers_review2008} and references therein), few studies exist on the coalescence dynamics of viscoelastic drops \citep{varma2020} [Fig.~\ref{fig:setup}(bc)]. Previous works \citep{Yue2005, zdravkov2003} exploring the effects of viscoelasticity on drop coalescence have focussed on film drainage as the two drops come together before coalescence. Here our focus is on the merging after initial contact between drops, which is mediated by the growth of a bridge that typically grows as $h \sim t^\alpha$ where $h$ is the bridge size and $t$ the time after contact \citep{hopper_1990, Eggers1999, Rocha2001, Wu2004, Yao2005, thoroddsen2005, Aarts2005, Ristenpart2006, Paulsen2011, HernandezSanchez2012, Eddi2013,  Xia2019}. The corresponding rate-of-deformation is estimated to diverge as $\dot h/h \sim 1/t$, and one thus anticipates a strongly elastic response during the initial phase of coalescence, when the flow time scale is much shorter than the relaxation time of polymers in the fluid -- much like in the case of bouncing putty. It has thus far remained elusive how polymers and the singularity at the moment of contact interact, and how this affects the merging process.

In this Letter, we resolve the coalescence dynamics of viscoelastic liquids by experiments on aqueous polymer solutions of varying concentration. We consider the coalescence of both freely suspended drops and of drops in contact with a substrate, cf. Fig.~\ref{fig:snapshots}. For both cases it is found that polymer stress dramatically changes the spatial structure of the bridge, as evident in Fig.~\ref{fig:setup}(bc). Yet, surprisingly, the temporal growth of the bridge is only mildly affected by the polymers. These features are explained and quantified by a similarity theory for sessile drops, revealing that viscoelastic coalescence is very different in nature compared to pinch-off.

\begin{figure}[tb!]%
	\centering\includegraphics{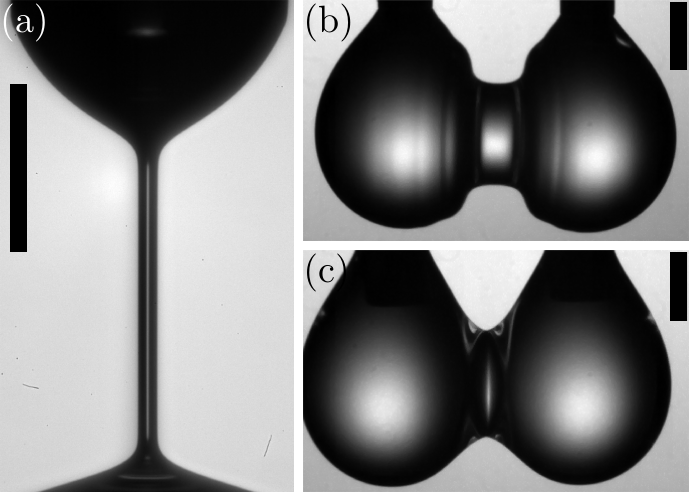}%
	\caption{Pinching and merging of viscoelastic polymer solutions. (a) Pinch-off of a 2.0 wt\% PEO drop gives rise to elongated threads. Coalescence of (b) pure water drops, and (c) 2.0 wt\% PEO drops. Polymer stretching inside the bridge markedly enhances the curvature of the connecting bridge. Scale bars indicate 500~$\mu$m.
	}%
	\label{fig:setup}%
\end{figure}%

\begin{figure}
\includegraphics{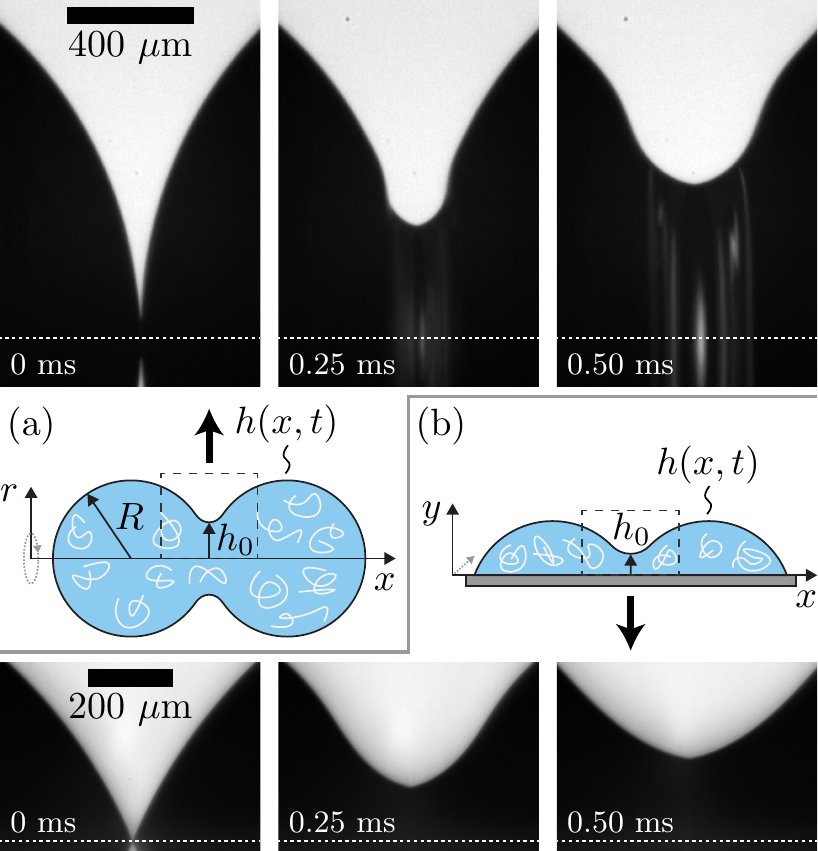}%
\caption{\label{fig:snapshots} 
Coalescence of viscoelastic drops (PEO solution with concentration 1.0 wt\% in two sister geometries: (a) ``spherical", consisting of two freely suspended drops, and (b) ``sessile", consisting of drops on a substrate. The three snapshots show a close-up of the bridge at three different times. The central bridge heigh $h_0(t)$ is defined with respect to the white dashed lines, indicating (a) $r = 0$, and (b) $y = 0$.}
\end{figure}

\paragraph{Experimental procedure}%
Coalescence experiments were performed using viscoelastic solutions of polyethylene oxide (PEO, $M_W = 4.0  \times 10^6$ g/mol, Sigma-Aldrich) in water (MilliQ, Millipore Corporation). Each solution was mixed with a magnetic stirrer for at least 24 hours, resulting in highly homogeneous solutions. While drop breakup is modified already for minute addition of polymer \citep{bonn2020, clasen2006}, elastic effects in coalescence require very high concentrations---up to 1.0 wt\% and 2.0 wt\%. We measured the shear viscosities ($\eta$) of the solutions using a rheometer (MCR 502 with CP50-1, Anton Paar), and the relaxation times ($\lambda$) using extensional thinning in a pendant drop geometry \citep{anna2001} [e.g., Fig.~\ref{fig:setup}(a)]. Calibrations are provided in the Supplementary Information~\cite{SI}. A ratio of time scales related to the material properties of the system can be defined as the material Deborah number ${\rm De_m}=\lambda/\tau$, where $\tau=\sqrt{\rho R^3/\gamma}\,  \approx 5.8$ ms is the inertio-capillary timescale based on the liquid density $\rho$, drop size $R$ and surface tension $\gamma$. In our experiments ${\rm De_m}$ ranges from $0.08$ to $10$. We note, however, that the relevant deformation rate during the early stages of coalescence scales as $1/t$, so that a more significant ratio is the ``local instantaneous Deborah number" $\lambda/t$, which reaches values up to $\mathcal{O}(10^4)$ at the smallest timescale that we can experimentally resolve. A strong polymer stretching and elastic effect is, therefore, anticipated during the early stages of coalescence. 

Typical snapshots of coalescence experiments are shown in Fig.~\ref{fig:snapshots} for  ``spherical" and ``sessile" geometries. In both cases, two symmetric drops are generated on needles (inner and outer diameters 0.52 mm and 0.82 mm, respectively) by a syringe pump (PHD 2000, Harvard Apparatus). They are brought into contact by very slowly increasing the drop volume, such that the approach velocity (in present experiments, three orders smaller than typical bridge velocity) can be neglected \cite{Eddi2013}. The coalescence dynamics is recorded by a high speed camera (Nova S12, Photron, with 12$\times$ zoom lens, Navitar), allowing for frame rates up to 200k fps, and resolutions down to 1 $\mu$m/pixel. For sessile drops, we use glass substrates (Menzel-Gl\"aser) that are made hydrophobic, and we use data with contact angles $45^\circ \leq\theta \leq 55^\circ$. We focus on the spatio-temporal evolution of the bridge shape, $h(x,t)$, as defined in Fig.~\ref{fig:snapshots}. This profile is extracted using a custom sub-pixel interface tracking code. Finding the initial time of coalescence from direct imaging is challenging, yet important to determine the coalescence exponent $\alpha$. Here $t=0$ is determined by extrapolating a power-law $h_0 \propto t^\alpha$, cf. SI \cite{SI}.

\paragraph{Time}%
The temporal evolution of the bridge growth is only mildly affected by the presence of polymers. This is evident from the data in Fig.~\ref{fig:time}. In panel (a) we show the growth of the minimal bridge radius $h_0(t)$ for spherical drops over the full range of polymer concentrations. The data nearly fall on top of one another, and closely follow the dynamics of pure water (included as dark blue symbols). A weak trend is observed with increasing polymer concentration, leading to slightly slower dynamics. Importantly, however, the exact same power-law growth  $h_0 \propto t^\alpha$ is found for all concentrations. The fitted values of $\alpha$, the coalescence exponent, are shown in panel (b): all polymer concentrations are consistent with $\alpha \approx 1/2$, which is the exponent for Newtonian coalescence of low-viscosity spherical drops \citep{Paulsen2011,Duchemin2003,Aarts2005}. 

\begin{figure}[tb!]%
	\centering\includegraphics[width=0.5\textwidth]{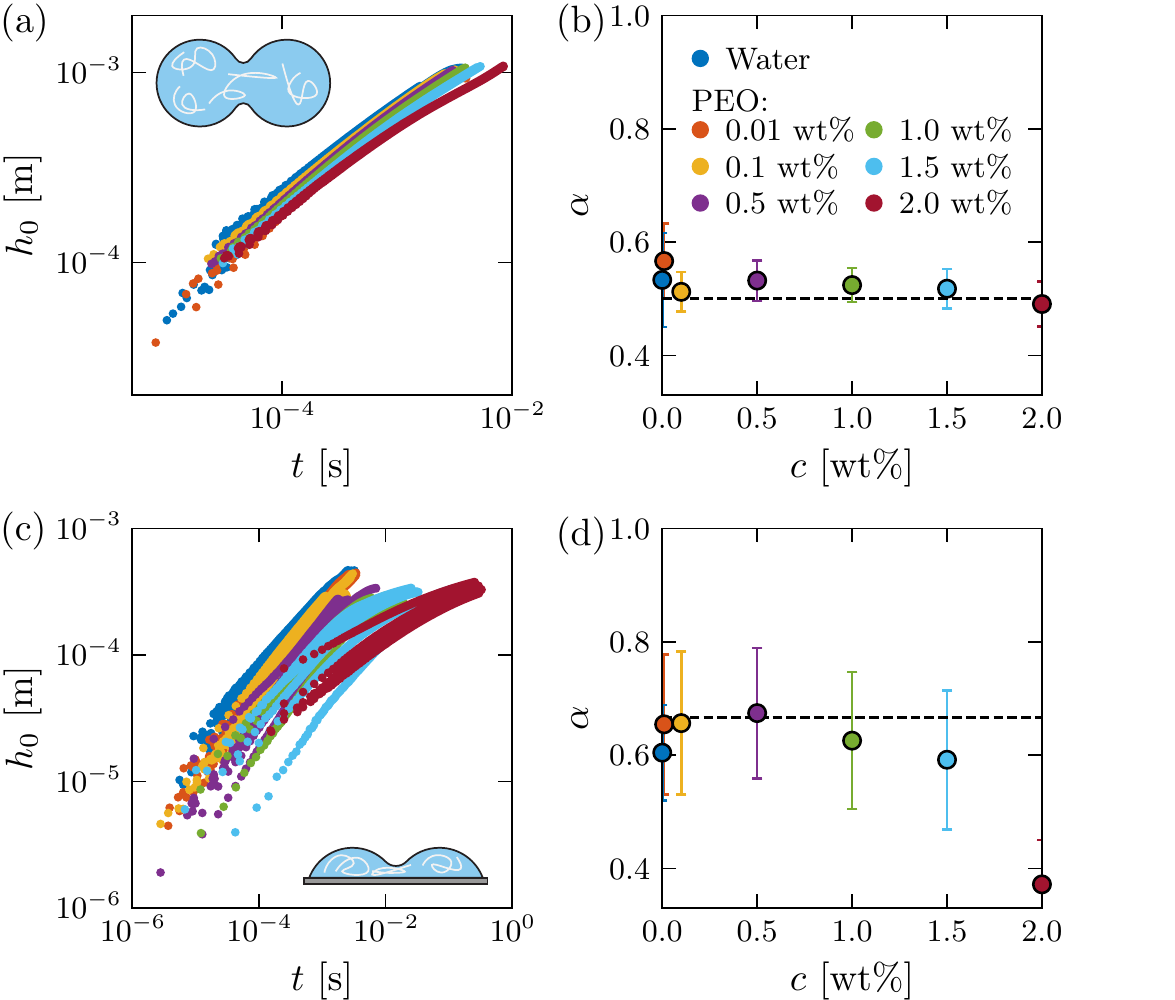}%
	\caption{Temporal coalescence dynamics. (a) Minimum bridge radius $h_0$ as a function of time for spherical drop coalescence, and the (b) fitted exponent $\alpha$; the dashed line indicates $\alpha = 1/2$. (c) Minimum bridge height $h_0$ as a function of time for sessile drop coalescence ($45^\circ \leq \theta \leq 55^\circ$), and the associated (d) fitted exponent $\alpha$; the dashed line indicates $\alpha = 2/3$. The error bars on $\alpha$ arise from a combination of the uncertainty in determining $t=0$, and the averaging different experiments with the same liquid.}%
	\label{fig:time}%
\end{figure}%

A similar behaviour is observed for sessile drops [Fig.~\ref{fig:time}(c,d)]. The data in panel (c) appear more scattered, which we attribute to the ``wetting" nature of these experiments, for it is known that the variability of the contact angle at the moment of coalescence affects the prefactor of the power-law growth \cite{Ristenpart2006,HernandezSanchez2012}. Once again, however, the coalescence exponent is close to that of pure water, which for sessile drops is $\alpha=2/3$ \citep{Eddi2013}. Only  the highest concentration exhibits a true departure from $2/3$, giving a lower exponent. 
Importantly, however,  up to concentrations of 1 wt\% there is no measurable effect of polymers on the coalescence exponent. We recall that a minute amount of polymer (down to 0.001 wt\%) \citep{bonn2020} is already sufficient to dramatically change the breakup of drops, from algebraic to exponential thinning. It is therefore truly remarkable that the coalescence exponent is completely unaffected by the presence of polymers at such high concentrations.

\paragraph{Space}%
Polymer stretching fundamentally alters the stress singularity during coalescence, and changes the
spatial structure of the bridge. Indeed, one observes a dramatic difference in bridge curvature in Figs.~\ref{fig:setup}(b,c). Laplace's law of capillarity implies that the enhanced curvature for PEO solutions is due to strong polymer stresses. This polymer stretching in fact leads to a breakdown of the self-similarity observed for the coalescence of pure water drops. Figure~\ref{fig:space} shows bridge profiles $h(x,t)$ at various times, scaled according to the Newtonian similarity solutions. The spherical and sessile cases have the same vertical scaling $h/h_0$, but call for a different scaling of the horizontal position, respectively, as $xR/h_0^2$ for spherical drops of radius $R$, and as $x \tan \theta/h_0$ for sessile drops with contact angle $\theta$ \citep{Eddi2013}. 

Figures~\ref{fig:space} (a,c) correspond to the reference cases of pure water. The scaled profiles exhibit a collapse, revealing the self-similar nature of Newtonian coalescence. Panels (b,d) report the corresponding data for PEO solutions, scaled in this Newtonian way. The selected data correspond to the highest concentrations for which we still observed the Newtonian coalescence exponents (2.0 wt\% for the spherical and 1.0 wt\% for the sessile case -- see figure \ref{fig:time}(b,d); for completeness, results for other concentrations are reported in SI \cite{SI}). The coalescence profiles for PEO no longer collapse with these scalings. The breakdown of Newtonian self-similarity emphasises the importance of polymeric stresses, which we now elucidate.

\begin{figure}[tb!]%
	\centering\includegraphics[width=0.5\textwidth]{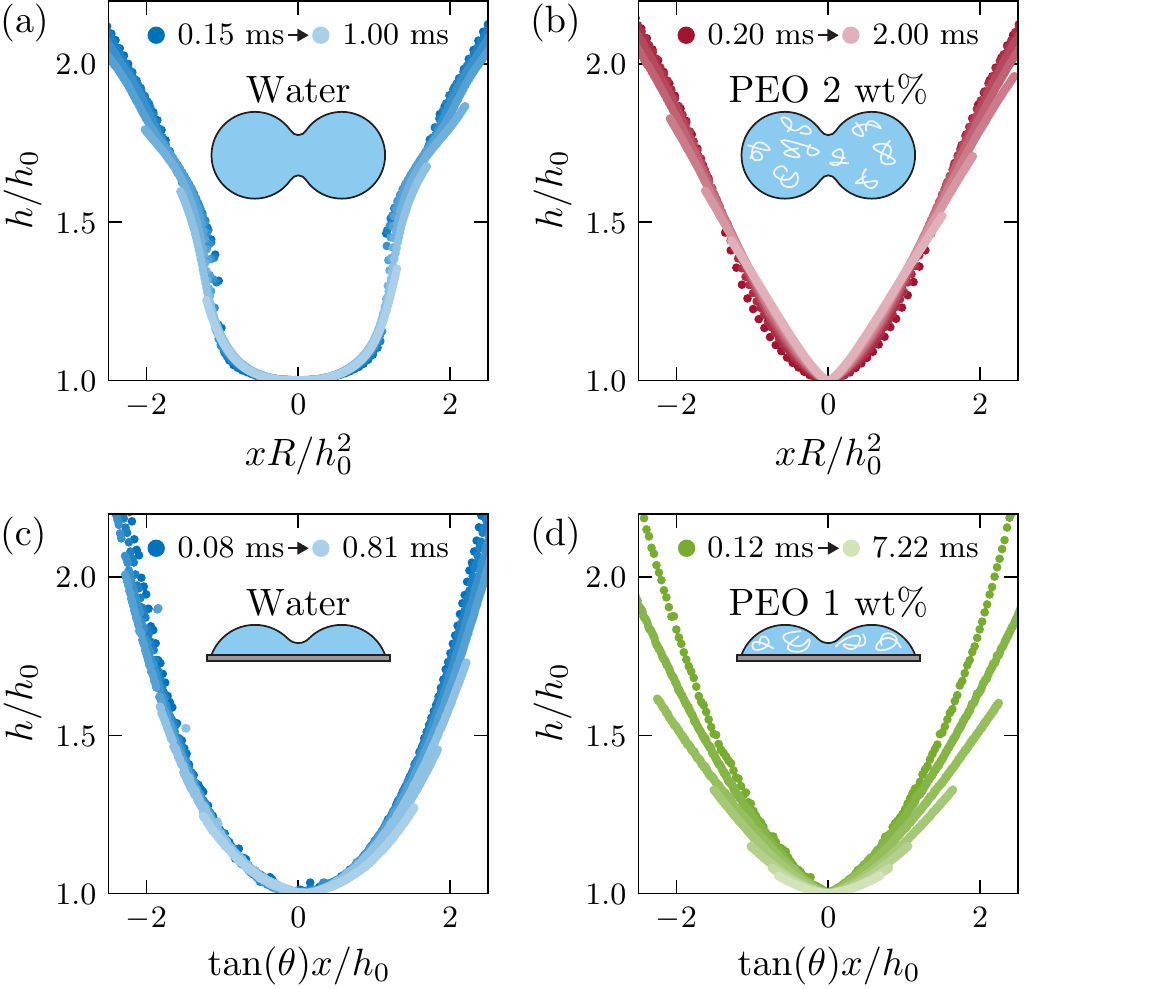}%
	\caption{Spatial coalescence dynamics, for spherical drops (a,b) and sessile drops (c,d). Bridge profiles at different times are rescaled according to the Newtonian scaling laws. For pure water, the rescaled bridge profiles collapse (a,c). A breakdown of this Newtonian self-similarity is found for PEO solutions (b,d). (Spherical: 2 wt\% PEO, $t/\lambda =0.003 \to 0.03$, sessile: 1 wt\% PEO, $t/\lambda =0.004 \to 0.25$.)}%
	\label{fig:space}%
\end{figure}%

\paragraph{Viscoelastic singularity}
An explicit analysis of the elastic singularity can be performed for sessile drops, where the geometry near coalescence is that of a wedge with angle $\theta$ as sketched in Fig.~\ref{fig:collapse}(a), and the ensuing flow structure is considerably simple \citep{Ristenpart2006, HernandezSanchez2012, Eddi2013}. The wedge geometry near the point of coalescence has only a single length scale $h_0$, and is more amenable to analytical treatment than a spherical one which involves a distinct, horizontal scale $h_0^2/R$. The wedge geometry for small $\theta$ offers a further simplification, which is exploited below. 

We start by noting that at the center of the bridge ($x=0$), the flow is nearly purely extensional in the vertical direction, with an extensional rate $\dot h_0/h_0 \sim 1/t$ that is very large at early times. The stress at $x=0$ at early times, thus, reads
\begin{equation}\label{eq:extension}
\sigma \sim \bar \eta_\infty \frac{\dot h_0}{h_0} \sim \frac{\bar \eta_\infty}{t}, 
\end{equation}
where $\bar{\eta}_{\infty}$ is the extensional viscosity at high rates \citep{book-bird, gauri1997extensional}. At such high stresses, nonlinear polymer relaxation becomes important and can be captured by constitutive relations such as the FENE-P or the Giesekus model. 
\begin{figure}[tb!]%
	\centering\includegraphics[width=0.5\textwidth]{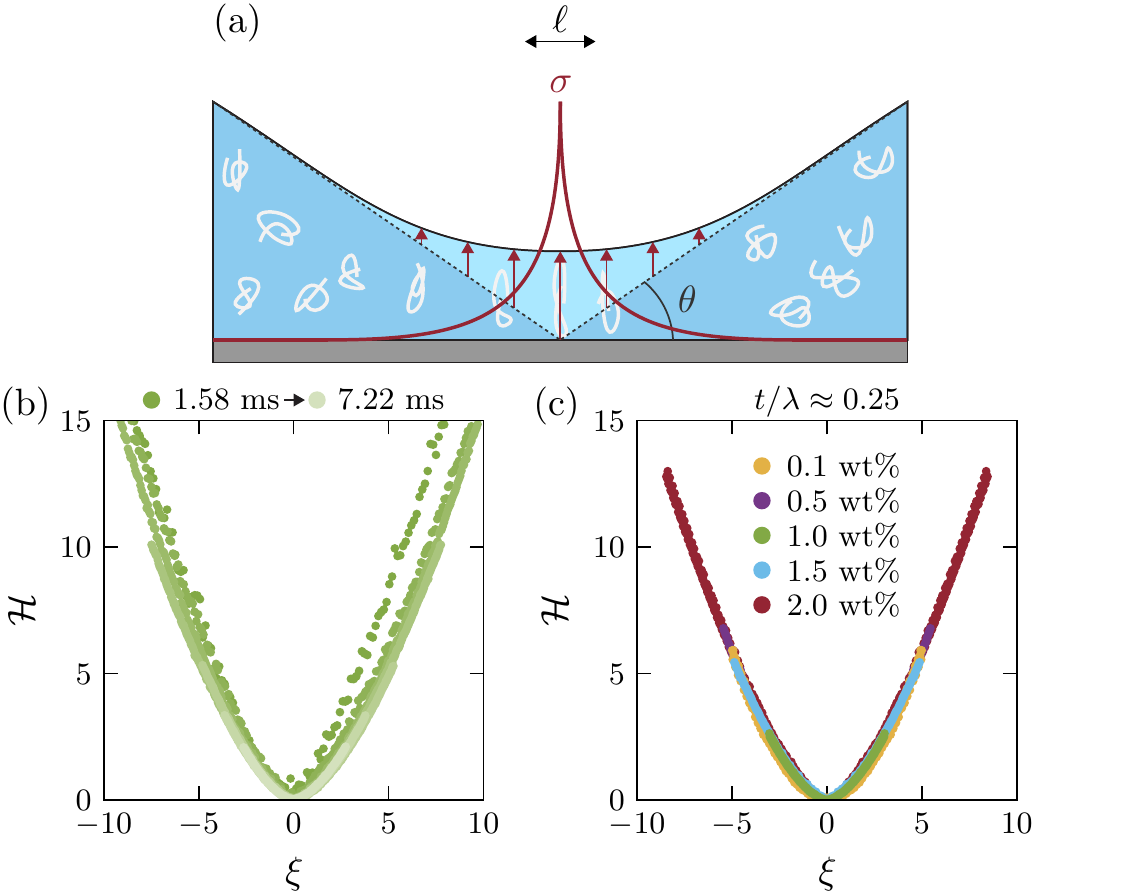}%
	\caption{Self-similarity of sessile viscoelastic coalescence. (a) Schematic of the elastic singularity in wedge coalescence due to polymer stretching in the vertical direction. The singular polymer stress (red line) decays over a characteristic distance $\ell$ that is small compared to the bridge width. (b,c) Bridge profiles scaled according to the viscoelastic self-similar prediction (\ref{eq:hsim}) for (b) different times (PEO 1.0 wt\%), and (c) different PEO concentrations at $t/\lambda\approx0.25$ (profiles symmetrized by averaging left-right). The data collapse on top of each other.}
	\label{fig:collapse}%
\end{figure}%

The amount of stretching will be much less pronounced away from $x=0$ [cf. Fig.~\ref{fig:collapse}(a)], and we wish to identify the characteristic distance $\ell$ over which the singular polymer stress (\ref{eq:extension}) decays. The central region of high polymer stretching is bordered by a region where polymer relaxation is negligible for $t/\lambda\ll 1$. This is a purely elastic region, where the polymer stretch can be found kinematically from the flow field \citep{eggers-2020-jfm}. Specifically, this involves comparing the height near $x= 0$ to the original height $\theta x$ prior to deformation. For small angles, the resulting polymer stress reads (cf. SI~\cite{SI})

\begin{equation}\label{eq:sigmasessile}
\sigma  \sim  \frac{Gh_0}{\theta x}, \quad \mathrm{for} \quad t \ll \lambda,
\end{equation}
where the appearance of $G$, the elastic modulus of the medium, underlines the purely elastic nature. The extent $x \approx \ell$ of the central bridge region can be found by matching stresses (\ref{eq:extension}) and (\ref{eq:sigmasessile}). This gives

\begin{equation}\label{eq:ell}
 \frac{G h_0 t}{\theta \bar \eta_\infty} \sim \frac{h_0 t}{\theta \lambda} \equiv \ell,
\end{equation}
where we have used $\bar \eta_\infty/G \sim \lambda$ to express the result in terms of the calibrated relaxation time. We note here that $\ell$ is much smaller, by a factor $t/\lambda$, than the horizontal coalescence scale $h_0/\theta$ for pure water. 

\paragraph{Viscoelastic self-similarity}
The emergence of length scale $\ell$ explains the breakdown of the self-similarity in Fig.~\ref{fig:space}, and provides a new horizontal scale for viscoelastic coalescence. However, we still need to identify the appropriate vertical scale. This is done by balancing the polymer stress, $\bar \eta_\infty/t$, to the capillary pressure, $\gamma h''$ in the bridge region, which leads us to 
\begin{equation}\label{eq:hsim}
h(x,t) - h_0  \sim  \frac{t h_0^2 G}{\lambda \gamma}\,  \mathcal H(\xi), \quad \mathrm{with} \quad \xi = \frac{x}{\ell} .
\end{equation}
This predicts a new self-similar regime where the bridge is described by a universal shape $\mathcal H(x/\ell)$. 

To test these scaling predictions, we accordingly rescale the experimental data. Figure ~\ref{fig:collapse}(b) shows the result for the 1.0 wt\% solution [same data as in Fig.~\ref{fig:space}(d)]. The data now exhibit a very good collapse which validates the emergence of the new, viscoelastic, self-similarity. We also compare the results from different polymer concentrations in Fig.~\ref{fig:collapse}(c), each taken at a dimensionless time $t/\lambda \approx 0.25$ (arbitrarily chosen, while keeping experimental resolution in mind). Since $G\lambda/\bar \eta_\infty$ in \eqref{eq:hsim} is expected to vary with polymer concentration, the scaling of vertical axis involves an adjustable parameter (reported in \cite{SI}), while the horizontal scale $\ell$ is left parameter-free. An excellent collapse can indeed be obtained, as seen in Fig.~\ref{fig:collapse}(c).

\paragraph{Discussion}%
Elasticity affects droplet coalescence in a remarkable fashion. While the spatial structure of the bridge is affected in a fundamental way, with sharp bridge profiles induced by polymer stretching, the temporal coalescence exponent remains unaffected. This second feature can be rationalised from our similarity analysis for sessile drops. The polymer stress in coalescence remains confined to a very narrow region, whose size depends on the ``local instantaneous Deborah number" $\lambda/t$. This narrow region turns out insufficient to alter the inertio-capillary coalescence exponent. This scenario is markedly different from droplet breakup for which polymer elasticity acts everywhere along the elongated filament. In breakup, the effective local Deborah number (relaxation time times the elongation rate) approaches a constant value along the entire thread \cite{clasen2006}, and therefore is able to dictate the temporal evolution.

As a future perspective, the next step would be to identify the structure in spherical drop coalescence}. The spherical case lacks a slender limit and is therefore more intricate, also due to two length scales in the problem already in the Newtonian case. It would be of interest to numerically investigate the initial phase of coalescence using different types of constitutive relations and different coalescence geometries. Our experiments, however, show that the scenario for sessile and spherical drops is qualitatively similar. These findings will be important for a plethora of applications involving merging of polymeric drops.

\begin{acknowledgments}
The authors thank J. Eggers for discussions. We acknowledge support from an Industrial Partnership Programme of NWO, co-financed by Oc{\'e}-Technologies B.V., University of Twente, and Eindhoven University of Technology. 
Further support from the European Union's Horizon 2020 research and innovation programme LubISS (No. 722497), NWO Vici (No. 680-47-63) is acknowledged. 
\end{acknowledgments}

\bibliography{visco_lens}

\begin{thebibliography}{37}%
\makeatletter
\providecommand \@ifxundefined [1]{%
 \@ifx{#1\undefined}
}%
\providecommand \@ifnum [1]{%
 \ifnum #1\expandafter \@firstoftwo
 \else \expandafter \@secondoftwo
 \fi
}%
\providecommand \@ifx [1]{%
 \ifx #1\expandafter \@firstoftwo
 \else \expandafter \@secondoftwo
 \fi
}%
\providecommand \natexlab [1]{#1}%
\providecommand \enquote  [1]{``#1''}%
\providecommand \bibnamefont  [1]{#1}%
\providecommand \bibfnamefont [1]{#1}%
\providecommand \citenamefont [1]{#1}%
\providecommand \href@noop [0]{\@secondoftwo}%
\providecommand \href [0]{\begingroup \@sanitize@url \@href}%
\providecommand \@href[1]{\@@startlink{#1}\@@href}%
\providecommand \@@href[1]{\endgroup#1\@@endlink}%
\providecommand \@sanitize@url [0]{\catcode `\\12\catcode `\$12\catcode
  `\&12\catcode `\#12\catcode `\^12\catcode `\_12\catcode `\%12\relax}%
\providecommand \@@startlink[1]{}%
\providecommand \@@endlink[0]{}%
\providecommand \url  [0]{\begingroup\@sanitize@url \@url }%
\providecommand \@url [1]{\endgroup\@href {#1}{\urlprefix }}%
\providecommand \urlprefix  [0]{URL }%
\providecommand \Eprint [0]{\href }%
\providecommand \doibase [0]{http://dx.doi.org/}%
\providecommand \selectlanguage [0]{\@gobble}%
\providecommand \bibinfo  [0]{\@secondoftwo}%
\providecommand \bibfield  [0]{\@secondoftwo}%
\providecommand \translation [1]{[#1]}%
\providecommand \BibitemOpen [0]{}%
\providecommand \bibitemStop [0]{}%
\providecommand \bibitemNoStop [0]{.\EOS\space}%
\providecommand \EOS [0]{\spacefactor3000\relax}%
\providecommand \BibitemShut  [1]{\csname bibitem#1\endcsname}%
\let\auto@bib@innerbib\@empty
\bibitem [{\citenamefont {Tanner}(2000)}]{tanner2000engineering}%
  \BibitemOpen
  \bibfield  {author} {\bibinfo {author} {\bibfnamefont {R.~I.}\ \bibnamefont
  {Tanner}},\ }\href@noop {} {\emph {\bibinfo {title} {Engineering
  rheology}}},\ Vol.~\bibinfo {volume} {52}\ (\bibinfo  {publisher} {OUP
  Oxford},\ \bibinfo {year} {2000})\BibitemShut {NoStop}%
\bibitem [{\citenamefont {Renardy}(2000)}]{renardy2000mathematical}%
  \BibitemOpen
  \bibfield  {author} {\bibinfo {author} {\bibfnamefont {M.}~\bibnamefont
  {Renardy}},\ }\href@noop {} {\emph {\bibinfo {title} {Mathematical analysis
  of viscoelastic flows}}}\ (\bibinfo  {publisher} {SIAM},\ \bibinfo {year}
  {2000})\BibitemShut {NoStop}%
\bibitem [{\citenamefont {Hinch}(1993)}]{hinch1993}%
  \BibitemOpen
  \bibfield  {author} {\bibinfo {author} {\bibfnamefont {E.}~\bibnamefont
  {Hinch}},\ }\href {\doibase https://doi.org/10.1016/0377-0257(93)80029-B}
  {\bibfield  {journal} {\bibinfo  {journal} {J. Non-Newton. Fluid Mech.}\
  }\textbf {\bibinfo {volume} {50}},\ \bibinfo {pages} {161} (\bibinfo {year}
  {1993})}\BibitemShut {NoStop}%
\bibitem [{\citenamefont {Renardy}(1993)}]{renardy1993}%
  \BibitemOpen
  \bibfield  {author} {\bibinfo {author} {\bibfnamefont {M.}~\bibnamefont
  {Renardy}},\ }\href {\doibase https://doi.org/10.1016/0377-0257(93)80027-9}
  {\bibfield  {journal} {\bibinfo  {journal} {J. Non-Newton. Fluid Mech.}\
  }\textbf {\bibinfo {volume} {50}},\ \bibinfo {pages} {127} (\bibinfo {year}
  {1993})}\BibitemShut {NoStop}%
\bibitem [{\citenamefont {Evans}\ and\ \citenamefont
  {Sibley}(2008)}]{evans2008}%
  \BibitemOpen
  \bibfield  {author} {\bibinfo {author} {\bibfnamefont {J.}~\bibnamefont
  {Evans}}\ and\ \bibinfo {author} {\bibfnamefont {D.}~\bibnamefont {Sibley}},\
  }\href {\doibase https://doi.org/10.1016/j.jnnfm.2007.11.006} {\bibfield
  {journal} {\bibinfo  {journal} {J. Non-Newtonian Fluid Mech.}\ }\textbf
  {\bibinfo {volume} {153}},\ \bibinfo {pages} {12} (\bibinfo {year}
  {2008})}\BibitemShut {NoStop}%
\bibitem [{\citenamefont {Astarita}\ and\ \citenamefont
  {Apuzzo}(1965)}]{astarita1965}%
  \BibitemOpen
  \bibfield  {author} {\bibinfo {author} {\bibfnamefont {G.}~\bibnamefont
  {Astarita}}\ and\ \bibinfo {author} {\bibfnamefont {G.}~\bibnamefont
  {Apuzzo}},\ }\href {\doibase https://doi.org/10.1002/aic.690110514}
  {\bibfield  {journal} {\bibinfo  {journal} {AIChE Journal}\ }\textbf
  {\bibinfo {volume} {11}},\ \bibinfo {pages} {815} (\bibinfo {year}
  {1965})}\BibitemShut {NoStop}%
\bibitem [{\citenamefont {Liu}\ \emph {et~al.}(1995)\citenamefont {Liu},
  \citenamefont {Liao},\ and\ \citenamefont {Joseph}}]{joseph1995}%
  \BibitemOpen
  \bibfield  {author} {\bibinfo {author} {\bibfnamefont {Y.~J.}\ \bibnamefont
  {Liu}}, \bibinfo {author} {\bibfnamefont {T.~Y.}\ \bibnamefont {Liao}}, \
  and\ \bibinfo {author} {\bibfnamefont {D.~D.}\ \bibnamefont {Joseph}},\
  }\href {\doibase 10.1017/S0022112095004447} {\bibfield  {journal} {\bibinfo
  {journal} {J. Fluid Mech.}\ }\textbf {\bibinfo {volume} {304}},\ \bibinfo
  {pages} {321–342} (\bibinfo {year} {1995})}\BibitemShut {NoStop}%
\bibitem [{\citenamefont {Zenit}\ and\ \citenamefont
  {Feng}(2018)}]{jimmy_annual2018}%
  \BibitemOpen
  \bibfield  {author} {\bibinfo {author} {\bibfnamefont {R.}~\bibnamefont
  {Zenit}}\ and\ \bibinfo {author} {\bibfnamefont {J.}~\bibnamefont {Feng}},\
  }\href {\doibase 10.1146/annurev-fluid-122316-045114} {\bibfield  {journal}
  {\bibinfo  {journal} {Annu. Rev. of Fluid Mech.}\ }\textbf {\bibinfo {volume}
  {50}},\ \bibinfo {pages} {505} (\bibinfo {year} {2018})}\BibitemShut
  {NoStop}%
\bibitem [{\citenamefont {Bazilevskii}\ \emph {et~al.}(1990)\citenamefont
  {Bazilevskii}, \citenamefont {Entov},\ and\ \citenamefont
  {Rozhkov}}]{bazilevskii1990}%
  \BibitemOpen
  \bibfield  {author} {\bibinfo {author} {\bibfnamefont {A.~V.}\ \bibnamefont
  {Bazilevskii}}, \bibinfo {author} {\bibfnamefont {V.~M.}\ \bibnamefont
  {Entov}}, \ and\ \bibinfo {author} {\bibfnamefont {A.~N.}\ \bibnamefont
  {Rozhkov}},\ }in\ \href@noop {} {\emph {\bibinfo {booktitle} {Proceedings of
  the {T}hird {E}uropean {R}heology {C}onference}}},\ \bibinfo {editor} {edited
  by\ \bibinfo {editor} {\bibfnamefont {D.~R.}\ \bibnamefont {Oliver}}}\
  (\bibinfo  {publisher} {Elsevier {A}pplied {S}cience},\ \bibinfo {year}
  {1990})\ p.~\bibinfo {pages} {41}\BibitemShut {NoStop}%
\bibitem [{\citenamefont {Entov}\ and\ \citenamefont
  {Hinch}(1997)}]{entov1997jnnfm}%
  \BibitemOpen
  \bibfield  {author} {\bibinfo {author} {\bibfnamefont {V.~M.}\ \bibnamefont
  {Entov}}\ and\ \bibinfo {author} {\bibfnamefont {E.~J.}\ \bibnamefont
  {Hinch}},\ }\href {\doibase https://doi.org/10.1016/S0377-0257(97)00022-0}
  {\bibfield  {journal} {\bibinfo  {journal} {J. Non-{N}ewtonian Fluid Mech.}\
  }\textbf {\bibinfo {volume} {72}},\ \bibinfo {pages} {31} (\bibinfo {year}
  {1997})}\BibitemShut {NoStop}%
\bibitem [{\citenamefont {Anna}\ and\ \citenamefont
  {Mc{K}inley}(2001)}]{anna2001}%
  \BibitemOpen
  \bibfield  {author} {\bibinfo {author} {\bibfnamefont {S.~L.}\ \bibnamefont
  {Anna}}\ and\ \bibinfo {author} {\bibfnamefont {G.~H.}\ \bibnamefont
  {Mc{K}inley}},\ }\href@noop {} {\bibfield  {journal} {\bibinfo  {journal} {J.
  Rheol.}\ }\textbf {\bibinfo {volume} {45}},\ \bibinfo {pages} {115} (\bibinfo
  {year} {2001})}\BibitemShut {NoStop}%
\bibitem [{\citenamefont {Amarouchene}\ \emph {et~al.}(2001)\citenamefont
  {Amarouchene}, \citenamefont {Bonn}, \citenamefont {Meunier},\ and\
  \citenamefont {Kellay}}]{amarouchene2001prl}%
  \BibitemOpen
  \bibfield  {author} {\bibinfo {author} {\bibfnamefont {Y.}~\bibnamefont
  {Amarouchene}}, \bibinfo {author} {\bibfnamefont {D.}~\bibnamefont {Bonn}},
  \bibinfo {author} {\bibfnamefont {J.}~\bibnamefont {Meunier}}, \ and\
  \bibinfo {author} {\bibfnamefont {H.}~\bibnamefont {Kellay}},\ }\href@noop {}
  {\bibfield  {journal} {\bibinfo  {journal} {Phys. Rev. Lett.}\ }\textbf
  {\bibinfo {volume} {86}},\ \bibinfo {pages} {3558} (\bibinfo {year}
  {2001})}\BibitemShut {NoStop}%
\bibitem [{\citenamefont {Clasen}\ \emph {et~al.}(2006)\citenamefont {Clasen},
  \citenamefont {Eggers}, \citenamefont {Fontelos}, \citenamefont {Li},\ and\
  \citenamefont {McKinley}}]{clasen2006}%
  \BibitemOpen
  \bibfield  {author} {\bibinfo {author} {\bibfnamefont {C.}~\bibnamefont
  {Clasen}}, \bibinfo {author} {\bibfnamefont {J.}~\bibnamefont {Eggers}},
  \bibinfo {author} {\bibfnamefont {M.~A.}\ \bibnamefont {Fontelos}}, \bibinfo
  {author} {\bibfnamefont {J.}~\bibnamefont {Li}}, \ and\ \bibinfo {author}
  {\bibfnamefont {G.~H.}\ \bibnamefont {McKinley}},\ }\href {\doibase
  10.1017/S0022112006009633} {\bibfield  {journal} {\bibinfo  {journal} {J.
  Fluid Mech.}\ }\textbf {\bibinfo {volume} {556}},\ \bibinfo {pages}
  {283–308} (\bibinfo {year} {2006})}\BibitemShut {NoStop}%
\bibitem [{\citenamefont {Eggers}\ \emph {et~al.}(2020)\citenamefont {Eggers},
  \citenamefont {Herrada},\ and\ \citenamefont {Snoeijer}}]{eggers-2020-jfm}%
  \BibitemOpen
  \bibfield  {author} {\bibinfo {author} {\bibfnamefont {J.}~\bibnamefont
  {Eggers}}, \bibinfo {author} {\bibfnamefont {M.~A.}\ \bibnamefont {Herrada}},
  \ and\ \bibinfo {author} {\bibfnamefont {J.~H.}\ \bibnamefont {Snoeijer}},\
  }\href@noop {} {\bibfield  {journal} {\bibinfo  {journal} {J. Fluid Mech.}\
  }\textbf {\bibinfo {volume} {887}},\ \bibinfo {pages} {A19} (\bibinfo {year}
  {2020})}\BibitemShut {NoStop}%
\bibitem [{\citenamefont {Deblais}\ \emph {et~al.}(2020)\citenamefont
  {Deblais}, \citenamefont {Herrada}, \citenamefont {Eggers},\ and\
  \citenamefont {Bonn}}]{bonn2020}%
  \BibitemOpen
  \bibfield  {author} {\bibinfo {author} {\bibfnamefont {A.}~\bibnamefont
  {Deblais}}, \bibinfo {author} {\bibfnamefont {M.~A.}\ \bibnamefont
  {Herrada}}, \bibinfo {author} {\bibfnamefont {J.}~\bibnamefont {Eggers}}, \
  and\ \bibinfo {author} {\bibfnamefont {D.}~\bibnamefont {Bonn}},\ }\href
  {\doibase 10.1017/jfm.2020.765} {\bibfield  {journal} {\bibinfo  {journal}
  {J. Fluid Mech.}\ }\textbf {\bibinfo {volume} {904}},\ \bibinfo {pages} {R2}
  (\bibinfo {year} {2020})}\BibitemShut {NoStop}%
\bibitem [{\citenamefont {Abkarian}\ and\ \citenamefont
  {Stone}(2020)}]{abkarian2020}%
  \BibitemOpen
  \bibfield  {author} {\bibinfo {author} {\bibfnamefont {M.}~\bibnamefont
  {Abkarian}}\ and\ \bibinfo {author} {\bibfnamefont {H.~A.}\ \bibnamefont
  {Stone}},\ }\href {\doibase 10.1103/PhysRevFluids.5.102301} {\bibfield
  {journal} {\bibinfo  {journal} {Phys. Rev. Fluids}\ }\textbf {\bibinfo
  {volume} {5}},\ \bibinfo {pages} {102301(R)} (\bibinfo {year}
  {2020})}\BibitemShut {NoStop}%
\bibitem [{\citenamefont {Renardy}(2004)}]{renardy2004self}%
  \BibitemOpen
  \bibfield  {author} {\bibinfo {author} {\bibfnamefont {M.}~\bibnamefont
  {Renardy}},\ }\href@noop {} {\bibfield  {journal} {\bibinfo  {journal}
  {Rheology Reviews}\ }\textbf {\bibinfo {volume} {2}},\ \bibinfo {pages} {171}
  (\bibinfo {year} {2004})}\BibitemShut {NoStop}%
\bibitem [{\citenamefont {Eggers}\ and\ \citenamefont
  {Villermaux}(2008)}]{eggers_review2008}%
  \BibitemOpen
  \bibfield  {author} {\bibinfo {author} {\bibfnamefont {J.}~\bibnamefont
  {Eggers}}\ and\ \bibinfo {author} {\bibfnamefont {E.}~\bibnamefont
  {Villermaux}},\ }\href@noop {} {\bibfield  {journal} {\bibinfo  {journal}
  {Rep. Prog. Phys.}\ }\textbf {\bibinfo {volume} {71}},\ \bibinfo {pages}
  {036601} (\bibinfo {year} {2008})}\BibitemShut {NoStop}%
\bibitem [{\citenamefont {Varma}\ \emph {et~al.}(2020)\citenamefont {Varma},
  \citenamefont {Saha}, \citenamefont {Mukherjee}, \citenamefont
  {Bandopadhyay}, \citenamefont {Kumar},\ and\ \citenamefont
  {Chakraborty}}]{varma2020}%
  \BibitemOpen
  \bibfield  {author} {\bibinfo {author} {\bibfnamefont {S.~C.}\ \bibnamefont
  {Varma}}, \bibinfo {author} {\bibfnamefont {A.}~\bibnamefont {Saha}},
  \bibinfo {author} {\bibfnamefont {S.}~\bibnamefont {Mukherjee}}, \bibinfo
  {author} {\bibfnamefont {A.}~\bibnamefont {Bandopadhyay}}, \bibinfo {author}
  {\bibfnamefont {A.}~\bibnamefont {Kumar}}, \ and\ \bibinfo {author}
  {\bibfnamefont {S.}~\bibnamefont {Chakraborty}},\ }\href {\doibase
  10.1039/D0SM01663B} {\bibfield  {journal} {\bibinfo  {journal} {Soft Matter}\
  }\textbf {\bibinfo {volume} {16}},\ \bibinfo {pages} {10921} (\bibinfo {year}
  {2020})}\BibitemShut {NoStop}%
\bibitem [{\citenamefont {Yue}\ \emph {et~al.}(2005)\citenamefont {Yue},
  \citenamefont {Feng}, \citenamefont {Liu},\ and\ \citenamefont
  {Shen}}]{Yue2005}%
  \BibitemOpen
  \bibfield  {author} {\bibinfo {author} {\bibfnamefont {P.}~\bibnamefont
  {Yue}}, \bibinfo {author} {\bibfnamefont {J.~J.}\ \bibnamefont {Feng}},
  \bibinfo {author} {\bibfnamefont {C.}~\bibnamefont {Liu}}, \ and\ \bibinfo
  {author} {\bibfnamefont {J.}~\bibnamefont {Shen}},\ }\href {\doibase
  https://doi.org/10.1016/j.jnnfm.2005.07.002} {\bibfield  {journal} {\bibinfo
  {journal} {J. Non-Newtonian Fluid Mech.}\ }\textbf {\bibinfo {volume}
  {129}},\ \bibinfo {pages} {163} (\bibinfo {year} {2005})}\BibitemShut
  {NoStop}%
\bibitem [{\citenamefont {Zdravkov}\ \emph {et~al.}(2003)\citenamefont
  {Zdravkov}, \citenamefont {Peters},\ and\ \citenamefont
  {Meijer}}]{zdravkov2003}%
  \BibitemOpen
  \bibfield  {author} {\bibinfo {author} {\bibfnamefont {A.}~\bibnamefont
  {Zdravkov}}, \bibinfo {author} {\bibfnamefont {G.}~\bibnamefont {Peters}}, \
  and\ \bibinfo {author} {\bibfnamefont {H.}~\bibnamefont {Meijer}},\ }\href
  {\doibase https://doi.org/10.1016/S0021-9797(03)00466-1} {\bibfield
  {journal} {\bibinfo  {journal} {J. Colloid Interface Science}\ }\textbf
  {\bibinfo {volume} {266}},\ \bibinfo {pages} {195} (\bibinfo {year}
  {2003})}\BibitemShut {NoStop}%
\bibitem [{\citenamefont {Hopper}(1990)}]{hopper_1990}%
  \BibitemOpen
  \bibfield  {author} {\bibinfo {author} {\bibfnamefont {R.~W.}\ \bibnamefont
  {Hopper}},\ }\href {\doibase 10.1017/S002211209000235X} {\bibfield  {journal}
  {\bibinfo  {journal} {J. Fluid Mech.}\ }\textbf {\bibinfo {volume} {213}},\
  \bibinfo {pages} {349–375} (\bibinfo {year} {1990})}\BibitemShut {NoStop}%
\bibitem [{\citenamefont {Eggers}\ \emph {et~al.}(1999)\citenamefont {Eggers},
  \citenamefont {Lister},\ and\ \citenamefont {Stone}}]{Eggers1999}%
  \BibitemOpen
  \bibfield  {author} {\bibinfo {author} {\bibfnamefont {J.}~\bibnamefont
  {Eggers}}, \bibinfo {author} {\bibfnamefont {J.~R.}\ \bibnamefont {Lister}},
  \ and\ \bibinfo {author} {\bibfnamefont {H.~A.}\ \bibnamefont {Stone}},\
  }\href {\doibase 10.1017/S002211209900662X} {\bibfield  {journal} {\bibinfo
  {journal} {J. Fluid Mech.}\ }\textbf {\bibinfo {volume} {401}},\ \bibinfo
  {pages} {293} (\bibinfo {year} {1999})}\BibitemShut {NoStop}%
\bibitem [{\citenamefont {Menchaca-Rocha}\ \emph {et~al.}(2001)\citenamefont
  {Menchaca-Rocha}, \citenamefont {Mart\'{\i}nez-D\'avalos}, \citenamefont
  {N\'u\~nez}, \citenamefont {Popinet},\ and\ \citenamefont
  {Zaleski}}]{Rocha2001}%
  \BibitemOpen
  \bibfield  {author} {\bibinfo {author} {\bibfnamefont {A.}~\bibnamefont
  {Menchaca-Rocha}}, \bibinfo {author} {\bibfnamefont {A.}~\bibnamefont
  {Mart\'{\i}nez-D\'avalos}}, \bibinfo {author} {\bibfnamefont
  {R.}~\bibnamefont {N\'u\~nez}}, \bibinfo {author} {\bibfnamefont
  {S.}~\bibnamefont {Popinet}}, \ and\ \bibinfo {author} {\bibfnamefont
  {S.}~\bibnamefont {Zaleski}},\ }\href {\doibase 10.1103/PhysRevE.63.046309}
  {\bibfield  {journal} {\bibinfo  {journal} {Phys. Rev. E}\ }\textbf {\bibinfo
  {volume} {63}},\ \bibinfo {pages} {046309} (\bibinfo {year}
  {2001})}\BibitemShut {NoStop}%
\bibitem [{\citenamefont {Wu}\ \emph {et~al.}(2004)\citenamefont {Wu},
  \citenamefont {Cubaud},\ and\ \citenamefont {Ho}}]{Wu2004}%
  \BibitemOpen
  \bibfield  {author} {\bibinfo {author} {\bibfnamefont {M.}~\bibnamefont
  {Wu}}, \bibinfo {author} {\bibfnamefont {T.}~\bibnamefont {Cubaud}}, \ and\
  \bibinfo {author} {\bibfnamefont {C.-M.}\ \bibnamefont {Ho}},\ }\href
  {\doibase 10.1063/1.1756928} {\bibfield  {journal} {\bibinfo  {journal}
  {Phys. Fluids}\ }\textbf {\bibinfo {volume} {16}},\ \bibinfo {pages} {L51}
  (\bibinfo {year} {2004})}\BibitemShut {NoStop}%
\bibitem [{\citenamefont {Yao}\ \emph {et~al.}(2005)\citenamefont {Yao},
  \citenamefont {Maris}, \citenamefont {Pennington},\ and\ \citenamefont
  {Seidel}}]{Yao2005}%
  \BibitemOpen
  \bibfield  {author} {\bibinfo {author} {\bibfnamefont {W.}~\bibnamefont
  {Yao}}, \bibinfo {author} {\bibfnamefont {H.~J.}\ \bibnamefont {Maris}},
  \bibinfo {author} {\bibfnamefont {P.}~\bibnamefont {Pennington}}, \ and\
  \bibinfo {author} {\bibfnamefont {G.~M.}\ \bibnamefont {Seidel}},\ }\href
  {\doibase 10.1103/PhysRevE.71.016309} {\bibfield  {journal} {\bibinfo
  {journal} {Phys. Rev. E}\ }\textbf {\bibinfo {volume} {71}},\ \bibinfo
  {pages} {016309} (\bibinfo {year} {2005})}\BibitemShut {NoStop}%
\bibitem [{\citenamefont {Thoroddsen}\ \emph {et~al.}(2005)\citenamefont
  {Thoroddsen}, \citenamefont {Takehara},\ and\ \citenamefont
  {Etoh}}]{thoroddsen2005}%
  \BibitemOpen
  \bibfield  {author} {\bibinfo {author} {\bibfnamefont {S.~T.}\ \bibnamefont
  {Thoroddsen}}, \bibinfo {author} {\bibfnamefont {K.}~\bibnamefont
  {Takehara}}, \ and\ \bibinfo {author} {\bibfnamefont {T.~G.}\ \bibnamefont
  {Etoh}},\ }\href {\doibase 10.1017/S0022112004003076} {\bibfield  {journal}
  {\bibinfo  {journal} {J. Fluid Mech.}\ }\textbf {\bibinfo {volume} {527}},\
  \bibinfo {pages} {85–114} (\bibinfo {year} {2005})}\BibitemShut {NoStop}%
\bibitem [{\citenamefont {Aarts}\ \emph {et~al.}(2005)\citenamefont {Aarts},
  \citenamefont {Lekkerkerker}, \citenamefont {Guo}, \citenamefont {Wegdam},\
  and\ \citenamefont {Bonn}}]{Aarts2005}%
  \BibitemOpen
  \bibfield  {author} {\bibinfo {author} {\bibfnamefont {D.~G. A.~L.}\
  \bibnamefont {Aarts}}, \bibinfo {author} {\bibfnamefont {H.~N.~W.}\
  \bibnamefont {Lekkerkerker}}, \bibinfo {author} {\bibfnamefont
  {H.}~\bibnamefont {Guo}}, \bibinfo {author} {\bibfnamefont {G.~H.}\
  \bibnamefont {Wegdam}}, \ and\ \bibinfo {author} {\bibfnamefont
  {D.}~\bibnamefont {Bonn}},\ }\href {\doibase 10.1103/PhysRevLett.95.164503}
  {\bibfield  {journal} {\bibinfo  {journal} {Phys. Rev. Lett.}\ }\textbf
  {\bibinfo {volume} {95}},\ \bibinfo {pages} {164503} (\bibinfo {year}
  {2005})}\BibitemShut {NoStop}%
\bibitem [{\citenamefont {Ristenpart}\ \emph {et~al.}(2006)\citenamefont
  {Ristenpart}, \citenamefont {McCalla}, \citenamefont {Roy},\ and\
  \citenamefont {Stone}}]{Ristenpart2006}%
  \BibitemOpen
  \bibfield  {author} {\bibinfo {author} {\bibfnamefont {W.~D.}\ \bibnamefont
  {Ristenpart}}, \bibinfo {author} {\bibfnamefont {P.~M.}\ \bibnamefont
  {McCalla}}, \bibinfo {author} {\bibfnamefont {R.~V.}\ \bibnamefont {Roy}}, \
  and\ \bibinfo {author} {\bibfnamefont {H.~A.}\ \bibnamefont {Stone}},\ }\href
  {\doibase 10.1103/PhysRevLett.97.064501} {\bibfield  {journal} {\bibinfo
  {journal} {Phys. Rev. Lett.}\ }\textbf {\bibinfo {volume} {97}},\ \bibinfo
  {pages} {064501} (\bibinfo {year} {2006})}\BibitemShut {NoStop}%
\bibitem [{\citenamefont {Paulsen}\ \emph {et~al.}(2011)\citenamefont
  {Paulsen}, \citenamefont {Burton},\ and\ \citenamefont
  {Nagel}}]{Paulsen2011}%
  \BibitemOpen
  \bibfield  {author} {\bibinfo {author} {\bibfnamefont {J.~D.}\ \bibnamefont
  {Paulsen}}, \bibinfo {author} {\bibfnamefont {J.~C.}\ \bibnamefont {Burton}},
  \ and\ \bibinfo {author} {\bibfnamefont {S.~R.}\ \bibnamefont {Nagel}},\
  }\href {\doibase 10.1103/PhysRevLett.106.114501} {\bibfield  {journal}
  {\bibinfo  {journal} {Phys. Rev. Lett.}\ }\textbf {\bibinfo {volume} {106}},\
  \bibinfo {pages} {114501} (\bibinfo {year} {2011})}\BibitemShut {NoStop}%
\bibitem [{\citenamefont {Hern\'andez-S\'anchez}\ \emph
  {et~al.}(2012)\citenamefont {Hern\'andez-S\'anchez}, \citenamefont {Lubbers},
  \citenamefont {Eddi},\ and\ \citenamefont {Snoeijer}}]{HernandezSanchez2012}%
  \BibitemOpen
  \bibfield  {author} {\bibinfo {author} {\bibfnamefont {J.~F.}\ \bibnamefont
  {Hern\'andez-S\'anchez}}, \bibinfo {author} {\bibfnamefont {L.~A.}\
  \bibnamefont {Lubbers}}, \bibinfo {author} {\bibfnamefont {A.}~\bibnamefont
  {Eddi}}, \ and\ \bibinfo {author} {\bibfnamefont {J.~H.}\ \bibnamefont
  {Snoeijer}},\ }\href {\doibase 10.1103/PhysRevLett.109.184502} {\bibfield
  {journal} {\bibinfo  {journal} {Phys.~Rev.~Lett.}\ }\textbf {\bibinfo
  {volume} {109}},\ \bibinfo {pages} {184502} (\bibinfo {year}
  {2012})}\BibitemShut {NoStop}%
\bibitem [{\citenamefont {Eddi}\ \emph {et~al.}(2013)\citenamefont {Eddi},
  \citenamefont {Winkels},\ and\ \citenamefont {Snoeijer}}]{Eddi2013}%
  \BibitemOpen
  \bibfield  {author} {\bibinfo {author} {\bibfnamefont {A.}~\bibnamefont
  {Eddi}}, \bibinfo {author} {\bibfnamefont {K.~G.}\ \bibnamefont {Winkels}}, \
  and\ \bibinfo {author} {\bibfnamefont {J.~H.}\ \bibnamefont {Snoeijer}},\
  }\href {\doibase 10.1103/PhysRevLett.111.144502} {\bibfield  {journal}
  {\bibinfo  {journal} {Phys.~Rev.~Lett.}\ }\textbf {\bibinfo {volume} {111}},\
  \bibinfo {pages} {144502} (\bibinfo {year} {2013})}\BibitemShut {NoStop}%
\bibitem [{\citenamefont {Xia}\ \emph {et~al.}(2019)\citenamefont {Xia},
  \citenamefont {He},\ and\ \citenamefont {Zhang}}]{Xia2019}%
  \BibitemOpen
  \bibfield  {author} {\bibinfo {author} {\bibfnamefont {X.}~\bibnamefont
  {Xia}}, \bibinfo {author} {\bibfnamefont {C.}~\bibnamefont {He}}, \ and\
  \bibinfo {author} {\bibfnamefont {P.}~\bibnamefont {Zhang}},\ }\href
  {\doibase 10.1073/pnas.1910711116} {\bibfield  {journal} {\bibinfo  {journal}
  {Proceedings of the National Academy of Sciences}\ }\textbf {\bibinfo
  {volume} {116}},\ \bibinfo {pages} {23467} (\bibinfo {year}
  {2019})}\BibitemShut {NoStop}%
\bibitem [{SI()}]{SI}%
  \BibitemOpen
  \href@noop {} {\bibinfo  {journal} {See Supplemental Material at [URL will be
  inserted by publisher] for calibrations and details on the slender analysis}\
  }\BibitemShut {NoStop}%
\bibitem [{\citenamefont {Duchemin}\ \emph {et~al.}(2003)\citenamefont
  {Duchemin}, \citenamefont {Eggers},\ and\ \citenamefont
  {Josserand}}]{Duchemin2003}%
  \BibitemOpen
\bibfield  {journal} {  }\bibfield  {author} {\bibinfo {author} {\bibfnamefont
  {L.}~\bibnamefont {Duchemin}}, \bibinfo {author} {\bibfnamefont
  {J.}~\bibnamefont {Eggers}}, \ and\ \bibinfo {author} {\bibfnamefont
  {C.}~\bibnamefont {Josserand}},\ }\href {\doibase 10.1017/S0022112003004646}
  {\bibfield  {journal} {\bibinfo  {journal} {J. Fluid Mech.}\ }\textbf
  {\bibinfo {volume} {487}},\ \bibinfo {pages} {167} (\bibinfo {year}
  {2003})}\BibitemShut {NoStop}%
\bibitem [{\citenamefont {Bird}\ \emph {et~al.}(1987)\citenamefont {Bird},
  \citenamefont {Curtiss}, \citenamefont {Armstrong},\ and\ \citenamefont
  {Hassager}}]{book-bird}%
  \BibitemOpen
  \bibfield  {author} {\bibinfo {author} {\bibfnamefont {R.~B.}\ \bibnamefont
  {Bird}}, \bibinfo {author} {\bibfnamefont {C.~F.}\ \bibnamefont {Curtiss}},
  \bibinfo {author} {\bibfnamefont {R.~C.}\ \bibnamefont {Armstrong}}, \ and\
  \bibinfo {author} {\bibfnamefont {O.}~\bibnamefont {Hassager}},\ }\href@noop
  {} {\emph {\bibinfo {title} {Dynamics of {P}olymeric {L}iquids}}}\ (\bibinfo
  {publisher} {John {W}iley \& {S}ons},\ \bibinfo {year} {1987})\BibitemShut
  {NoStop}%
\bibitem [{\citenamefont {Gauri}\ and\ \citenamefont
  {Koelling}(1997)}]{gauri1997extensional}%
  \BibitemOpen
  \bibfield  {author} {\bibinfo {author} {\bibfnamefont {V.}~\bibnamefont
  {Gauri}}\ and\ \bibinfo {author} {\bibfnamefont {K.~W.}\ \bibnamefont
  {Koelling}},\ }\href {\doibase 10.1007/BF00368133} {\bibfield  {journal}
  {\bibinfo  {journal} {Rheol. Acta}\ }\textbf {\bibinfo {volume} {36}},\
  \bibinfo {pages} {555} (\bibinfo {year} {1997})}\BibitemShut {NoStop}%
\end{thebibliography}%


\begin{thebibliography}{4}%
\makeatletter
\providecommand \@ifxundefined [1]{%
 \@ifx{#1\undefined}
}%
\providecommand \@ifnum [1]{%
 \ifnum #1\expandafter \@firstoftwo
 \else \expandafter \@secondoftwo
 \fi
}%
\providecommand \@ifx [1]{%
 \ifx #1\expandafter \@firstoftwo
 \else \expandafter \@secondoftwo
 \fi
}%
\providecommand \natexlab [1]{#1}%
\providecommand \enquote  [1]{``#1''}%
\providecommand \bibnamefont  [1]{#1}%
\providecommand \bibfnamefont [1]{#1}%
\providecommand \citenamefont [1]{#1}%
\providecommand \href@noop [0]{\@secondoftwo}%
\providecommand \href [0]{\begingroup \@sanitize@url \@href}%
\providecommand \@href[1]{\@@startlink{#1}\@@href}%
\providecommand \@@href[1]{\endgroup#1\@@endlink}%
\providecommand \@sanitize@url [0]{\catcode `\\12\catcode `\$12\catcode
  `\&12\catcode `\#12\catcode `\^12\catcode `\_12\catcode `\%12\relax}%
\providecommand \@@startlink[1]{}%
\providecommand \@@endlink[0]{}%
\providecommand \url  [0]{\begingroup\@sanitize@url \@url }%
\providecommand \@url [1]{\endgroup\@href {#1}{\urlprefix }}%
\providecommand \urlprefix  [0]{URL }%
\providecommand \Eprint [0]{\href }%
\providecommand \doibase [0]{http://dx.doi.org/}%
\providecommand \selectlanguage [0]{\@gobble}%
\providecommand \bibinfo  [0]{\@secondoftwo}%
\providecommand \bibfield  [0]{\@secondoftwo}%
\providecommand \translation [1]{[#1]}%
\providecommand \BibitemOpen [0]{}%
\providecommand \bibitemStop [0]{}%
\providecommand \bibitemNoStop [0]{.\EOS\space}%
\providecommand \EOS [0]{\spacefactor3000\relax}%
\providecommand \BibitemShut  [1]{\csname bibitem#1\endcsname}%
\let\auto@bib@innerbib\@empty
\bibitem [{\citenamefont {Stauffer}(1965)}]{Stauffer1965}%
  \BibitemOpen
  \bibfield  {author} {\bibinfo {author} {\bibfnamefont {C.~E.}\ \bibnamefont
  {Stauffer}},\ }\href {\doibase 10.1021/j100890a024} {\bibfield  {journal}
  {\bibinfo  {journal} {J. Phys. Chem.}\ }\textbf {\bibinfo {volume} {69}},\
  \bibinfo {pages} {1933} (\bibinfo {year} {1965})}\BibitemShut {NoStop}%
\bibitem [{\citenamefont {Anna}\ and\ \citenamefont
  {Mc{K}inley}(2001)}]{anna2001}%
  \BibitemOpen
  \bibfield  {author} {\bibinfo {author} {\bibfnamefont {S.~L.}\ \bibnamefont
  {Anna}}\ and\ \bibinfo {author} {\bibfnamefont {G.~H.}\ \bibnamefont
  {Mc{K}inley}},\ }\href@noop {} {\bibfield  {journal} {\bibinfo  {journal} {J.
  Rheol.}\ }\textbf {\bibinfo {volume} {45}},\ \bibinfo {pages} {115} (\bibinfo
  {year} {2001})}\BibitemShut {NoStop}%
\bibitem [{\citenamefont {Eggers}\ \emph {et~al.}(2020)\citenamefont {Eggers},
  \citenamefont {Herrada},\ and\ \citenamefont {Snoeijer}}]{eggers-2020-jfm}%
  \BibitemOpen
  \bibfield  {author} {\bibinfo {author} {\bibfnamefont {J.}~\bibnamefont
  {Eggers}}, \bibinfo {author} {\bibfnamefont {M.~A.}\ \bibnamefont {Herrada}},
  \ and\ \bibinfo {author} {\bibfnamefont {J.~H.}\ \bibnamefont {Snoeijer}},\
  }\href@noop {} {\bibfield  {journal} {\bibinfo  {journal} {J. Fluid Mech.}\
  }\textbf {\bibinfo {volume} {887}},\ \bibinfo {pages} {A19} (\bibinfo {year}
  {2020})}\BibitemShut {NoStop}%
\bibitem [{\citenamefont {Eggers}\ and\ \citenamefont
  {Fontelos}(2015)}]{Eggers2015}%
  \BibitemOpen
  \bibfield  {author} {\bibinfo {author} {\bibfnamefont {J.}~\bibnamefont
  {Eggers}}\ and\ \bibinfo {author} {\bibfnamefont {M.~A.}\ \bibnamefont
  {Fontelos}},\ }\href {\doibase 10.1017/CBO9781316161692} {\emph {\bibinfo
  {title} {Singularities: Formation, Structure, and Propagation}}}\ (\bibinfo
  {publisher} {Cambridge University Press},\ \bibinfo {year}
  {2015})\BibitemShut {NoStop}%
\end{thebibliography}%

\end{document}


\title{Supplementary Material for `When Elasticity Affects Drop Coalescence'}

\author{Pim J. Dekker}
\author{Michiel A. Hack}
\author{Walter Tewes}
\author{Charu Datt}
\author{Ambre Bouillant}
\author{Jacco H. Snoeijer}

\affiliation{Physics of Fluids Group, Mesa+ Institute, University of Twente, 7500 AE Enschede, The Netherlands}

\begin{abstract}
This Supplementary Material presents characterisations of the polymer solutions, complementary experimental data, and details on the scaling argument in the purely elastic regime.
\end{abstract}

\date{\today}

\maketitle

\section{Rheological characterization}
\begin{figure}[H]%
	\centering\includegraphics{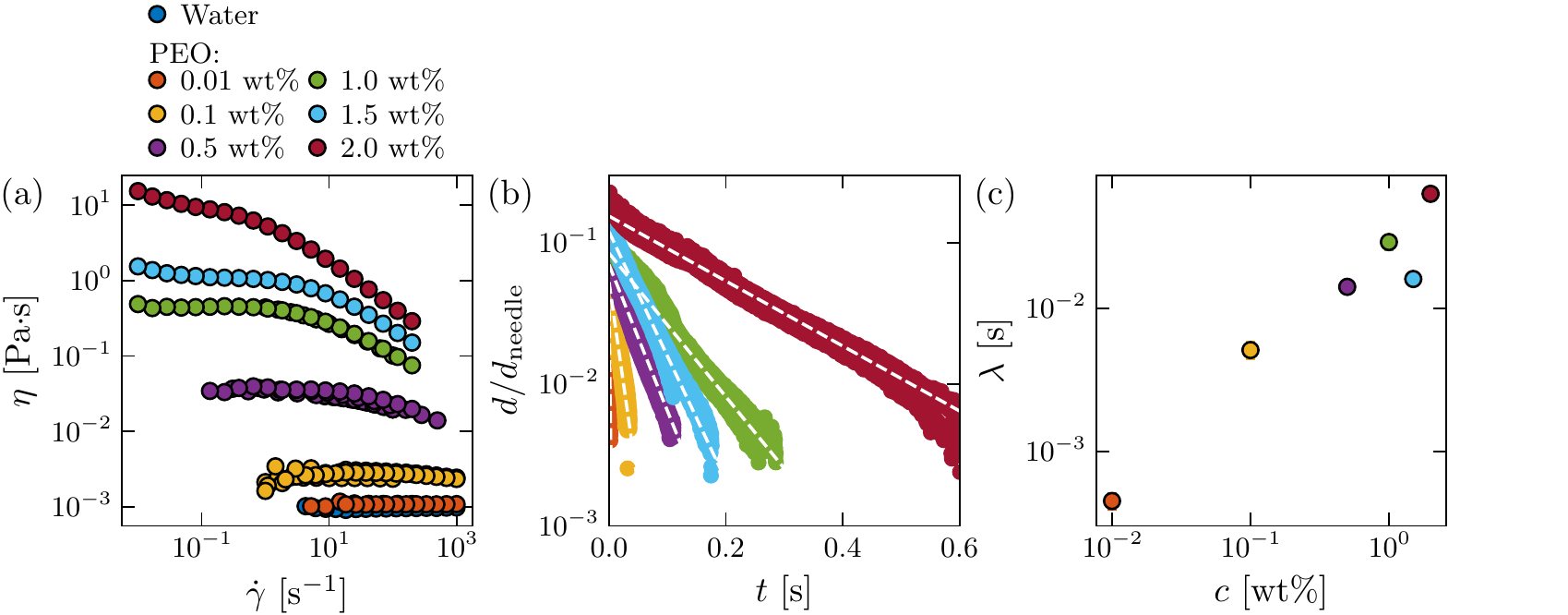}%
	\caption{(a) Viscosity $\eta$ as a function of shear rate $\dot{\gamma}$ for various PEO concentrations. Data were averaged for a few seconds for the high shear rates to a few minutes for the low shear rates.(b) Thinning dynamics of a thread for various PEO solutions. The thread has minimum diameter $d$. At least five experiments are shown for each solution. The white dashed lines show fits of the type $d/d_\mathrm{needle} \sim \exp(-t/3\lambda)$. (c) Relaxation time $\lambda$ as a function of PEO concentration $c$, as determined from the data in (b).
	}%
	\label{fig:suppfigrheometry}%
\end{figure}

A stock solution of polyethylene oxide (PEO) with $c=2$ wt$\%$ was prepared by weighing and mixing PEO with molecular weight $M_W = 4 \times 10^6$ g/mol (Sigma-Aldrich).  Lower concentrations down to 0.01 wt$\%$ were achieved by dilution. The physical properties of the fluid were characterized using two methods. The shear viscosity $\eta$ was measured using a rheometer (MCR 502 with CP50-1$^{\circ}$ cone-plate geometry spindle, Anton Paar) and the resulting values are shown in Fig.~\ref{fig:suppfigrheometry}(a) as a function of the shear rate $\dot{\gamma}$. Viscosity measurements were complemented by pendant drop experiments, which provide values for the surface tension $\gamma$ as well as the relaxation time $\lambda$ of the solutions. The surface tension, deduced by fitting the drop shape as in \citep{Stauffer1965},  is found to be insensitive to the polymer concentration:  $\gamma = 65\pm2$ mN/m. The relaxation time $\lambda$ was measured by studying the thread thinning dynamics \cite{anna2001}.  An example snapshot of a thread is shown in Fig.~1a of the main text. We performed at least five measurements per solution to extract the thread minimum diameter $d(t)$. We show in Fig.~\ref{fig:suppfigrheometry}b the thread time evolution on a semi-log scale, which is fitted with the function $d/d_\mathrm{needle} \sim \exp(-t/3\lambda)$,  where $d_\mathrm{needle}$ denotes the dispensing needle diameter and $t$ is the time. We provide in Fig.~\ref{fig:suppfigrheometry}c the deduced average relaxation times $\lambda$ as a function of the PEO concentration.

\newpage
\section{Finding the initial moment of drop contact}

\begin{figure}[h!]%
	\centering\includegraphics{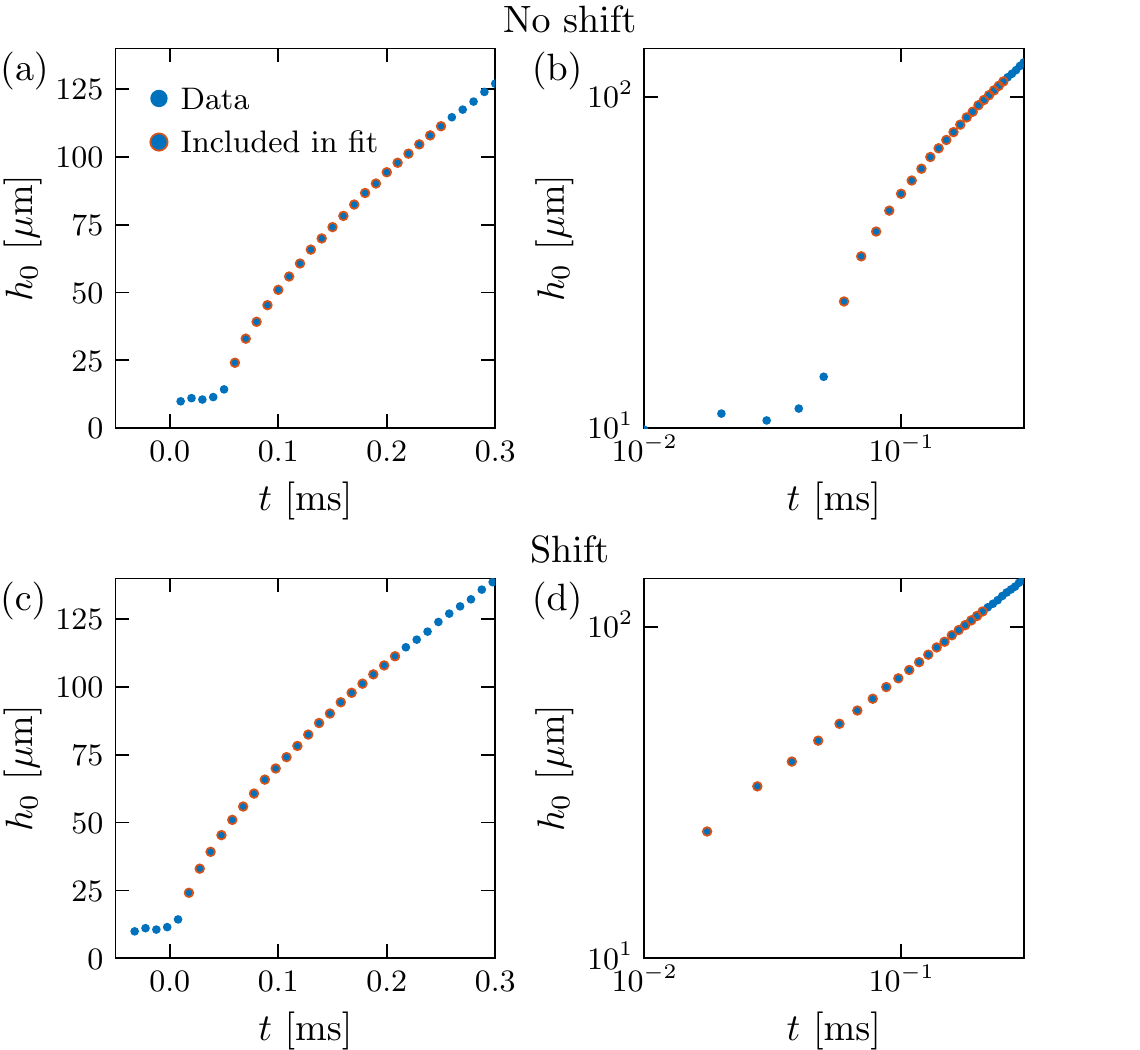}%
	\caption{Minimum bridge height $h_0$ as a function of time, with misdefined $t_0$ on (a) linear axes, and (b) loglog axes. The same data is shown with corrected $t_0$ on (c) linear axes, and (d) loglog axes.
	}%
	\label{fig:suppfigt0fit}%
\end{figure}

The accurate fitting of the exponent $\alpha$, defined as $h_0\propto(t-t_0)^\alpha$, relies on a good determination of $t_0$, i.e., the moment of first contact between the coalescing drops. A slightly misdefined $t_0$ leads to curved data on a log-log plot, affecting the fitted $\alpha$. Typical coalescence data for the sessile case obtained from optical microscopy is shown in Fig.~\ref{fig:suppfigt0fit}(a)-(b). The exact moment of first contact is difficult to determine from the experimental images---pixel resolution limits the visibility of small bridges, the moment of first contact can be in between two frames, and sideview images can lead to apparent contact before the actual contact. The method by which we determine the precise moment of $t_0$ is illustrated by the example measurement shown in Fig.~\ref{fig:suppfigt0fit}. We manually selected a range, starting at a frame where the coalescence had clearly started and ending at sufficiently small $t$ such that the growth is expected to follow a power law. For various small time shifts, we then checked if the selected data fits to a power law $h_0\propto(t-t_0)^\alpha$, by fitting (least squares method) a straight line where $\alpha$ was a free parameter. The value of $t_0$ was then determined by selecting the (small) time shift which resulted in the smallest fit residue. Fig.~\ref{fig:suppfigt0fit}(c)-(d) show the result of our method for the data shown in Fig.~\ref{fig:suppfigt0fit}(a)-(b). This procedure leaves $\alpha$ as a free parameter, which is reported in the main text. 

\newpage
\section{Additional data for various PEO concentrations}

\begin{figure}[H]%
	\centering\includegraphics[scale=0.75]{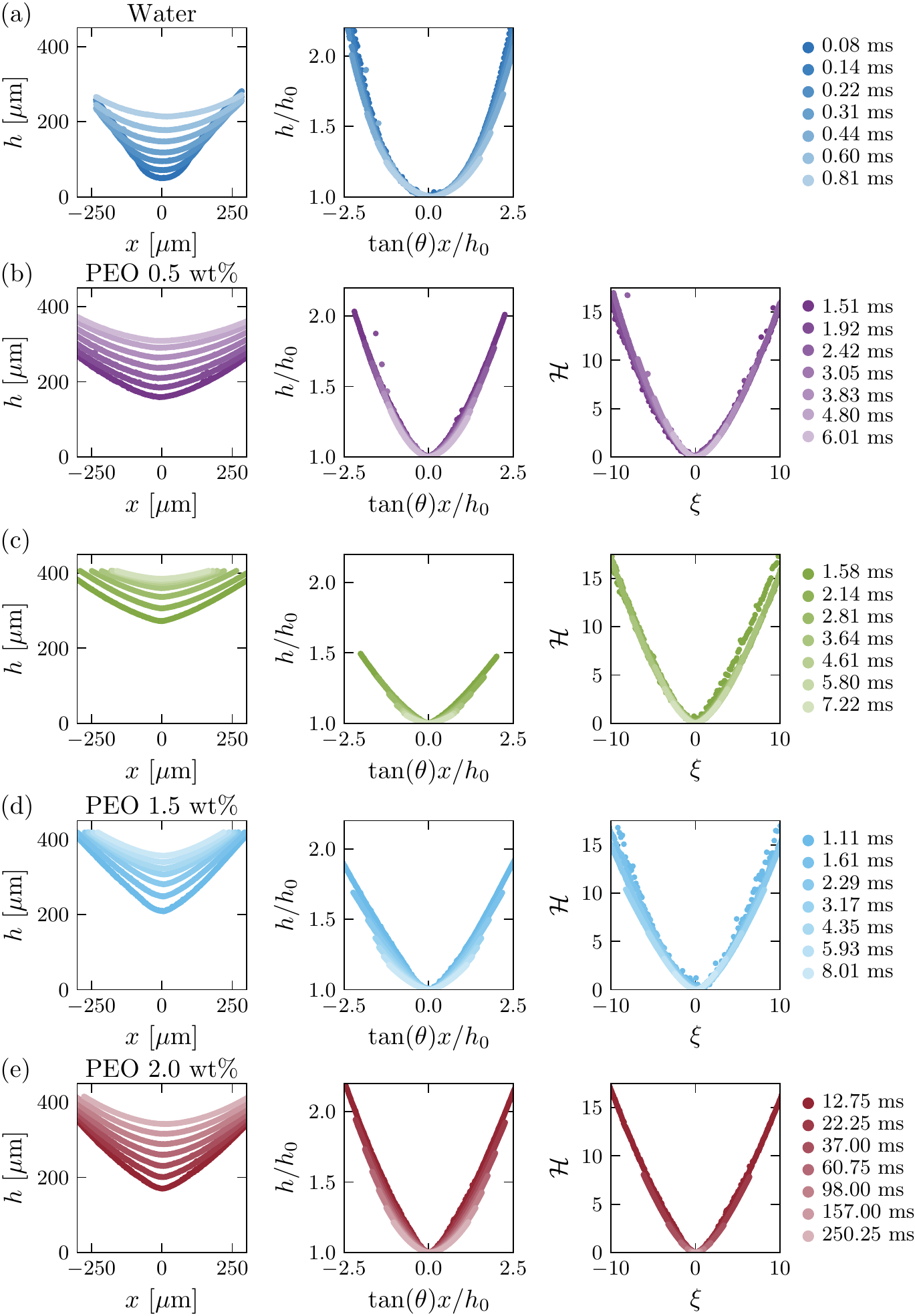}%
	\caption{Raw data and scaled data for sessile coalescence for various additional PEO concentrations. The leftmost column shows raw data, the middle column shows data scaled according to the Newtonian similarity scaling, the rightmost column shows the data scaled according the viscoelastic similarity scaling.  The Newtonian collapse progressively deteriorates with increasing PEO concentration. }%
	\label{fig:suppfigsessile}%
\end{figure}

\begin{figure}[H]%
	\centering\includegraphics[scale=0.75]{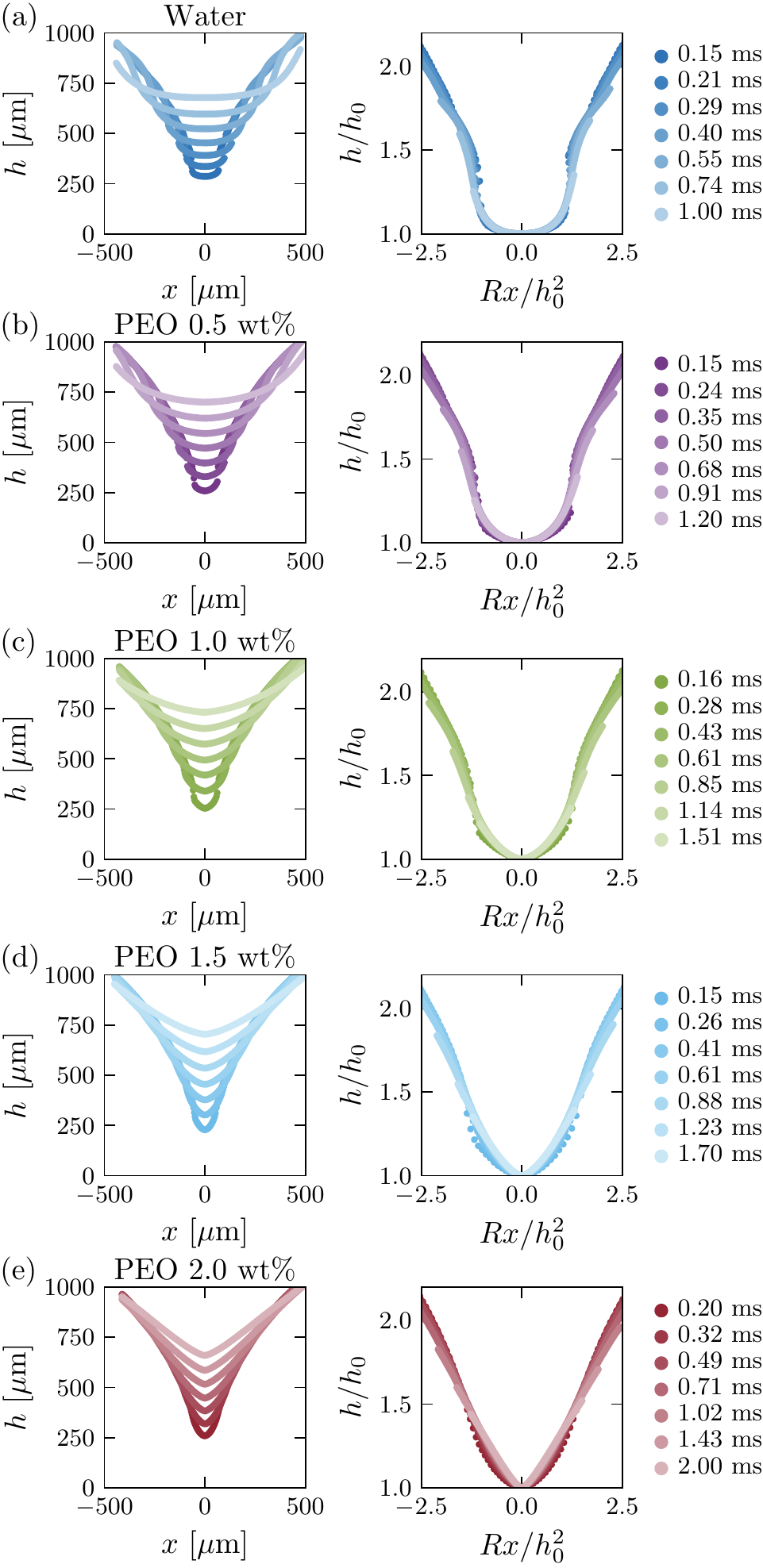}%
	\caption{Raw data and scaled data for spherical coalescence for various additional PEO concentrations. The left column shows raw data, the right column shows data scaled according to the Newtonian similarity scaling.  The Newtonian collapse progressively deteriorates with increasing PEO concentration.
	}%
	\label{fig:suppfigspherical}%
\end{figure}

In the main text, we present bridge profiles scaled by the Newtonian self-similar rescaling (Figure 4) and by the newly proposed viscoelastic self-similar rescaling (Figure 5). The main text shows profiles for selected concentrations and selected times. For completeness, here we provide the full set of data (scaled and unscaled) for all concentrations at various times. Figure~\ref{fig:suppfigsessile} gives the data for the sessile drop coalescence.  Figure~\ref{fig:suppfigspherical} gives the data for the spherical drop coalescence. Note how the Newtonian collapse progressively deteriorates with increasing PEO concentration.

\begin{figure}[H]%
	\centering\includegraphics{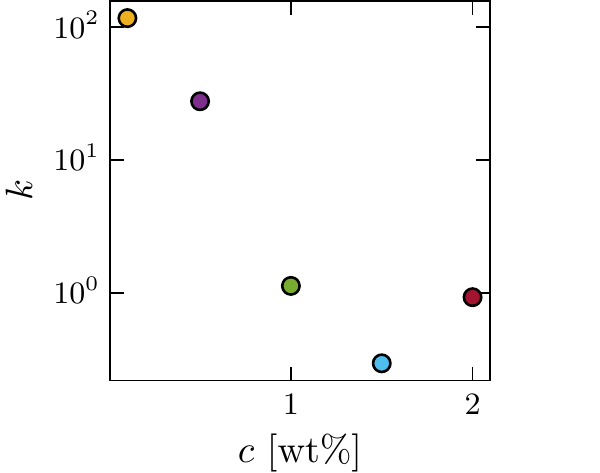}%
	\caption{The collapse of profiles in figure 5c involved a prefactor to the scaling law, equation (4) of the main text. The plot reports this prefactor $k$ for various concentrations. 
	}%
	\label{fig:refactor}%
\end{figure}%

\newpage
\section{The wedge geometry and the purely elastic regime}

\begin{figure}[H]%
	\centering\includegraphics[width = 0.5\textwidth]{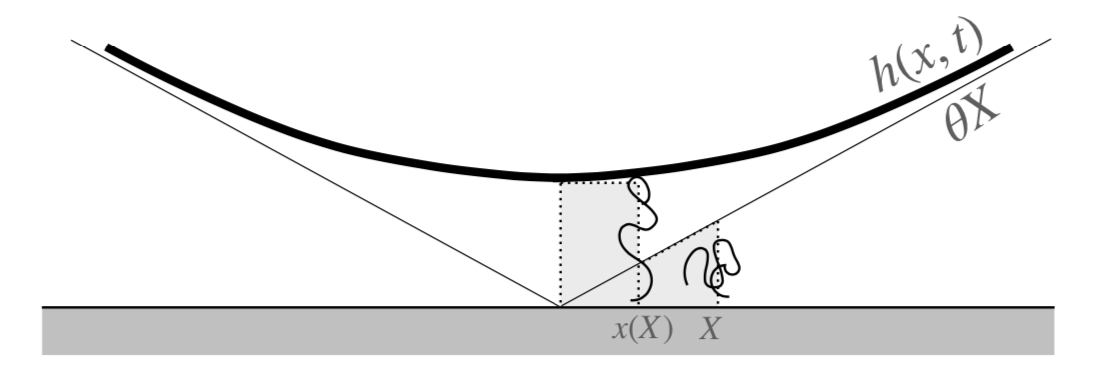}%
	\caption{Schematic of sessile lens coalescence in the bridge region.
	}%
	\label{fig:suppfigrwedge}%
\end{figure}

Here we motivate in more detail the scaling law in the purely elastic regime presented in equation (2) of the main text. For a stress-free initial condition, the polymer stress without relaxation reads \citep{eggers-2020-jfm}

\begin{equation}
\boldsymbol \sigma^p = G ~ \mathbf F \cdot \mathbf F^T,
\end{equation} 
where $\mathbf{F}$ is the deformation gradient tensor $\mathbf{F} = \dfrac{\partial \mathbf{x}}{\partial \mathbf{X}}$, and $\mathbf{x}$ and $\mathbf{X}$ are the Eulerian and Lagrangian coordinates, respectively (see figure \ref{fig:suppfigrwedge}). In the absence of polymer relaxation, the components of $\mathbf A = \mathbf F \cdot \mathbf F^T$ along principle directions give the square of the principle stretch that is induced by the flow. 

In the sessile drop coalescence, the predominant stretch near the center of the bridge is in the vertical direction. We thus need to evaluate $A_{yy}$. Furthermore, in the slender limit \citep{Eggers2015} and ignoring a thin viscous boundary layer near the wall, the flow will be approximately homogeneous across the film thickness. The stretching can then be evaluated by comparing $h(x,t)$ to the reference height $\theta X$ before the start of the coalescence, leading to 
 \begin{equation}
A_{yy} \approx \left(\dfrac{h(x,t)}{\theta X}\right)^2 \approx \left(\dfrac{h_0 (t)}{\theta X}\right)^2 .
\end{equation}
The second relation comes from an evaluation near the point of coalescence. Mass conservation in the two shaded regions in figure \ref{fig:suppfigrwedge} gives
\begin{equation}
\dfrac{1}{2}{X^2 \theta} = \int_0^x h(x,t)\ dx \approx h_0 (t) x. 
\end{equation} 
From the two relations one obtains $A_{yy} \sim h_0/ \theta x$. Therefore, the corresponding stress component $\sigma_{yy}  \sim G h_0/ \theta x$. This is equation (2) of the main text. 

\bibliography{visco_lens}